\documentclass[twocolumn,showpacs,preprintnumbers]{revtex4}
\usepackage{amssymb}
\usepackage{amsmath}
\usepackage{graphicx}
\usepackage{dcolumn}
\usepackage{bm}

\begin{document}

\title{Global Entanglement for Multipartite Quantum States}
\author{Chang-shui Yu}
\author{He-shan Song}
\email{hssong@dlut.edu.cn}
\affiliation{Department of Physics, Dalian University of Technology,\\
Dalian 116024, China}
\date{\today }

\begin{abstract}
Based on the residual entanglement [9] (Phys. Rev. A \textbf{71}, 044301
(2005)), we present the global entanglement for a multipartite quantum
state. The measure is shown to be also obtained by the bipartite partitions
of the multipartite state. The distinct characteristic of the global
entanglement is that it consists of the sum of different entanglement
contributions. The measure can provide sufficient and necessary condition of
fully separability for pure states and be conveniently extended to mixed
states by minimizing the convex hull. To test the sufficiency of the measure
for mixed states, we evaluate the global entanglement of bound entangled
states. The properties of the measure discussed finally show the global
entanglement is an entanglement monotone.
\end{abstract}

\maketitle

\section{Introduction}

Entanglement is an essential ingredient in the broad field of quantum
information theory. Quantification of entanglement as a central problem in
quantum information theory has attracted a lot of attention in recent years.

A lot of methods to quantify entanglement have been proposed. For bipartite
pure states, the partial entropy of the density matrix of a system defined
by
\begin{equation}
E(\psi )=-Tr(\rho \log _{2}\rho )=-\underset{i}{\sum }(\lambda _{i}\log
_{2}\lambda _{i}),
\end{equation}%
can provide a good measure of entanglement, where $\rho $ is the reduced
density matrix obtained by trace over one of the subsystems, $\lambda _{i}$
is the $i$th eigenvalue of $\rho $. For mixed states, the entanglement of
formation is defined by
\begin{equation}
E_{f}(\rho )=\min \underset{i}{\sum }p_{i}E(\psi _{i})
\end{equation}%
with $E(\psi _{i})$ the entanglement measure for the pure state $\psi _{i}$
corresponding to all the possible decompositions $\rho =\underset{i}{\sum }%
p_{i}\left\vert \psi _{i}\right\rangle \left\langle \psi _{i}\right\vert $.
Eq.(2) is a complex constrained optimization in mathematics, hence it is
hard to compute for a general mixed states, even numerically. Fortunately,
Wootters [1] has shown the remarkable concurrence for a bipartite system of
two spin (or pseudo spin) half particles, which can be well employed to
estimate bipartite entanglement and sheds new light on the quantification of
entanglement. Later, Armin Uhlmann [2] generalized the concurrence, and
Koenraad Audenaert et al [3] defined a concurrence vector for bipartite
systems in arbitrary dimension the length of which for pure states is proved
by Wootters [4] to be equal to the I-concurrence introduced by Rungta et al
[5], whilst they presented an effective method to extend the definition to
the case of mixed states by minimizing the convex hull.

There has also been ongoing efforts to investigate entanglement measure for
multipartite states [6-12]. In Ref. [6-9], the authors were focused on the
genuine multipartite entanglement measure which embodies a collective
property of multipartite systems. In Ref. [10], the authors extracted only
the correlation between two subsystem among a given multipartite system. The
authors in Ref. [11,12] only considered bipartite correlations between
single subsystems and the remainder, whilst they present the concept of
global entanglement because examples there showed that the entanglement
measure does not vanish for semiseparable states. However, the measure is
only confined to the pure quantum systems of qubits.

In this paper, starting with the residual entanglement [6,9], we present a
quantity which is shown to consist of the sum of different entanglement
contributions, and can also be considered as the sum of different
correlations, unlike the entanglement measures given in Ref. [6-12]. In this
sense, we also call it the global entanglement. The global entanglement can
provide a sufficient and necessary condition of full separability for pure
states. Furthermore, the global entanglement happens to be conveniently
obtained by the idea of bipartite partition of a multipartite quantum
states, which includes the case introduced in Ref. [11,12] (For tripartite
systems of qubits, our global entanglement is equivalent to the original
one). Extending the measure to mixed states by minimizing the convex hull,
we obtain the lower bound analogous to those in Ref. [3] and Ref. [9]. As an
application to test fully separability, we discuss the global entanglement
of three bound entangled states introduced in [13,14,15], respectively. The
properties of the global entanglement discussed finally shows that it is an
entanglement monotone. The paper is organized as follows: In section II,
firstly, we present the global entanglement for multipartite pure states;
secondly, we extend it to the case of mixed states; lastly, we discuss the
properties of the measure. The conclusion is drawn in section III.

\section{Global entanglement for multipartite quantum states}

\subsection{Global entanglement for multipartite pure states}

At first, let us recall the definition of the concurrence vectors [3, 16]
for bipartite states. Considering a bipartite pure state defined in $%
n_{1}\times n_{2}$ dimension, written by
\begin{equation*}
\left\vert \psi \right\rangle
_{AB}=\sum_{i=0}^{n_{1}-1}\sum_{j=0}^{n_{2}-1}a_{ij}\left\vert
i\right\rangle _{A}\otimes \left\vert j\right\rangle _{B},
\end{equation*}%
with $\sum_{i=0}^{n_{1}-1}\sum_{j=0}^{n_{2}-1}\left\vert a_{ij}\right\vert
^{2}=1.$ The concurrence vector \textbf{C }can be defined by%
\begin{equation*}
\mathbf{C}=(C_{00},C_{01},\cdot \cdot \cdot ,C_{\frac{n_{1}(n_{1}-1)}{2}%
\frac{n_{2}(n_{2}-1)}{2}}),
\end{equation*}%
where $C_{\alpha \beta }=\left\langle \psi _{AB}\right\vert s_{\alpha \beta
}\left\vert \psi _{AB}^{\ast }\right\rangle $ with $s_{\alpha \beta
}=L_{\alpha }\otimes L_{\beta }$; $L_{\alpha }$, $\alpha =1,$ $2$, $\cdot
\cdot \cdot ,$ $\frac{n_{1}(n_{1}-1)}{2}$ and $L_{\beta }$, $\beta
=1,2,\cdot \cdot \cdot ,\frac{n_{2}(n_{2}-1)}{2}$ are the generators of $%
SO(n_{1})$ and $SO(n_{2})$, respectively. The length of the vector $\mathbf{C%
}$ (we call it concurrence throughout the paper) hence be given by
\begin{equation*}
C(\left\vert \psi \right\rangle _{AB})=\sqrt{\sum_{\alpha \beta }\left\vert
C_{\alpha \beta }\right\vert ^{2}}.
\end{equation*}%
$C(\left\vert \psi \right\rangle _{AB})$ is an entanglement measure for
bipartite pure states. In particular, $C(\left\vert \psi \right\rangle _{AB})
$ can be reduced to Wootters' concurrence for $n_{1}=n_{2}=2$. For
multipartite quantum states, considering the bipartite partitions, the
multipartite states can be considered as bipartite quantum states.
Therefore, by the definition above, one can also obtain the concurrence of
such bipartite states.

For any bipartite mixed state $\rho
_{AB}=\sum\limits_{k=1}^{K}\omega _{k}\left\vert \psi
_{AB}^{k}\right\rangle \left\langle \psi _{AB}^{k}\right\vert $
defined in $n_{1}\times n_{2}$ dimension, the concurrece $C(\rho
_{AB})$ [9] can be given by
\begin{equation*}
C(\rho _{AB})=\underset{z\in C^{\alpha \beta }}{max}\lambda
_{1}(z)-\sum\limits_{i>1}\lambda _{i}(z).
\end{equation*}%
Here $\lambda _{j}(z)$ are the singular values of $\underset{\alpha =1}{%
\overset{n_{1}(n_{1}-1)/2}{\sum }\text{ }}\overset{n_{2}(n_{2}-1)/2}{%
\underset{\beta =1}{\sum }}z_{\alpha \beta }A_{\alpha \beta }$ in
decreasing order where $A_{\alpha \beta }=M^{1/2}\Phi ^{T}S_{\alpha
\beta }\Phi M^{1/2}$ with $\rho _{AB}=\Phi M\Phi ^{\dag }$ \ being
the eigenvalue decomposition, and $z_{\alpha \beta }=y_{\alpha \beta
}e^{i\phi _{\alpha
\beta }}$ are optimal parameters with $y_{\alpha \beta }>0$, $%
\sum\limits_{\alpha \beta }y_{\alpha \beta }^{2}=1$.

For tripartite pure quantum states of qubits, based on Wootters'
concurrence, the authors in Ref. [6] have defined the residual entanglement
(or 3 tangle) given by%
\begin{equation}
\tau _{ABC}=C_{A(BC)}^{2}-C_{AB}^{2}-C_{AC}^{2},
\end{equation}%
where $C_{AB}$ and $C_{AC}$ are the concurrences of the original pure state $%
\rho _{ABC}$ with traces taken over qubits $C$ and $B$, respectively. $%
C_{A(BC)}$ is the concurrence of $\rho _{A(BC)}$ with qubits $B$ and $C$
regarded as a single object. $\tau _{ABC}$ is shown to be the genuine
tripartite entanglement. As an extension of $\tau _{ABC}$ (3 tangle),
hyperdeterminant in Ref. [17] has been shown to be an entanglement monotone
and represent the genuine multipartite entanglement. However, it is easy to
find that the hyperdeterminant for higher dimensional systems and
multipartite system can not be explictly given conveniently. In particular,
so far the hyperdeterminant as an entanglement measure has not been able to
be extended to mixed systems. Therefore, the hyperdeterminant is difficult
to find the connection with the distribution of multipartite entanglement.
As an extension of eq. (3) or the distribution of entanglement, by
considering the concurrence of bipartite states in arbitrary dimension, Ref.
[9] has generalized eq. (3) to the higher-dimensional systems, multipartite
systems and mixed systems. However, it is unfortunate that so far no one has
been able to show whether the residual entanglements obtained in Ref. [9]
corresponding to different foci [6,9] are equal or not. Hence, strictly
speaking, the generalized residual entanglement $\tau $ can not be called
the exact genuine multipartite entanglement measure before $\tau $s are
proved to be equal. However, we can confirm that $\tau $s are relevant
quantities to genuine multipartite entanglement no matter whether they are
equal or not. For simplification, we call $\tau $ $n$-tangle corresponding
to the $n$ subscripts of $\tau $, e.g. $\tau _{1234}$ is called $4$-tangle.

For convenience and without loss of the generality, we first consider a
4-partite pure state $\Psi _{ABCD}$. According to Ref. [9], one can obtain
the following 10 equalities corresponding to different foci,%
\begin{equation}
C_{A(BCD)}^{2}=C_{A(BC)}^{2}+C_{AD}^{2}+\tau _{A\left( BC\right) D},
\end{equation}%
\begin{equation}
C_{B(ACD)}^{2}=C_{B(AC)}^{2}+C_{BD}^{2}+\tau _{B\left( AC\right) D},
\end{equation}%
\begin{equation}
C_{C(ABD)}^{2}=C_{C(AB)}^{2}+C_{CD}^{2}+\tau _{C\left( AB\right) D},
\end{equation}%
\begin{equation}
C_{D(BCA)}^{2}=C_{D(BC)}^{2}+C_{DA}^{2}+\tau _{D\left( BC\right) A},
\end{equation}%
\begin{equation}
C_{\left( AB\right) \left( CD\right) }^{2}=C_{\left( AB\right)
C}^{2}+C_{\left( AB\right) D}^{2}+\tau _{\left( AB\right) CD},
\end{equation}%
\begin{equation}
C_{\left( AC\right) \left( BD\right) }^{2}=C_{\left( AC\right)
B}^{2}+C_{\left( AC\right) D}^{2}+\tau _{\left( AC\right) BD},
\end{equation}%
\begin{equation}
C_{\left( AD\right) \left( BC\right) }^{2}=C_{\left( AD\right)
B}^{2}+C_{\left( AD\right) C}^{2}+\tau _{\left( AD\right) BC},
\end{equation}%
\begin{equation}
C_{\left( BC\right) \left( AD\right) }^{2}=C_{\left( BC\right)
A}^{2}+C_{\left( BC\right) D}^{2}+\tau _{\left( BC\right) AD},
\end{equation}%
\begin{equation}
C_{\left( BD\right) \left( AC\right) }^{2}=C_{\left( BD\right)
A}^{2}+C_{\left( BD\right) C}^{2}+\tau _{\left( BD\right) AC},
\end{equation}%
\begin{equation}
C_{\left( CD\right) \left( AB\right) }^{2}=C_{\left( CD\right)
A}^{2}+C_{\left( CD\right) B}^{2}+\tau _{\left( CD\right) AB},
\end{equation}%
where the brackets in the subscripts denote single objects and $C_{A(BC)}$
and $C_{AD}$ denote concurrences of the mixed state $\rho _{A\left(
BC\right) }$ and $\rho _{AD}$ which are obtained by tracing over qudits the
lost indices correspond to from $\Psi _{ABCD}$; the other analogous
notations in equations.(4-13) are defined in the similar way. It is worth
noting that the permutations of the qudits in a bracket do not change the
value of the left hand side of the equations. But the forms of the right
hand side of the former four equations (4-7) will be changed. Hence
considering all permutations of qudits in the former four equations, there
exist two other analogous equations [18] corresponding to each of them. As
given in Ref. [9], the analogous equation to eq. (3) has been shown to hold
for mixed states. That is to say, for any mixed state $\rho _{abc}$,%
\begin{equation*}
\tau _{abc}=C_{a(bc)}^{2}-C_{ab}^{2}-C_{ac}^{2},
\end{equation*}%
where $C_{a(bc)}$ is the concurrence of the mixed state $\rho _{a(bc)}$, $%
C_{ab}$ is the concurrence of the mixed state $\rho _{ab}$ which is given by
tracing over qudit $c$, $C_{ac}$ is defined analogously to $C_{ab}$.
According to the equation for tripartite mixed quantum systems given above,
we can expand above equations (4-13). For example, for eq. (4), we have
\begin{equation}
C_{A(BCD)}^{2}=C_{AB}^{2}+C_{AC}^{2}+\tau _{ABC}+C_{AD}^{2}+\tau _{A\left(
BC\right) D}.
\end{equation}%
The others are analogous. Summing all the equations up, one can obtain that
\begin{eqnarray}
&&C_{A(BCD)}^{2}+C_{B(ACD)}^{2}+C_{C(ABD)}^{2}+C_{D(BCA)}^{2} \\
&&+C_{\left( AB\right) (CD)}^{2}+C_{\left( AC\right) (BD)}^{2}+C_{\left(
AD\right) (BC)}^{2}  \notag \\
&&+C_{\left( BC\right) (AD)}^{2}+C_{\left( BD\right) (AC)}^{2}+C_{\left(
CD\right) (AB)}^{2}  \notag \\
&=&3\sum_{m,n\in S}C_{mn}^{2}+\frac{2}{3}\sum_{m,n,p\in S}\tau _{m\left(
np\right) }  \notag \\
&&+\frac{1}{6}\sum_{m,n,p,q\in S}\tau _{m\left( np\right) q}+\frac{1}{4}%
\sum_{m,n,p,q\in S}\tau _{\left( mn\right) pq},  \notag
\end{eqnarray}%
where $S=\{A,B,C,D\},$ $C_{mn}$ denote the concurrence vector of the reduced
state $\rho _{mn}$, $\tau _{m\left( np\right) }$ are the 3-tangle of the
reduced state $\rho _{m\left( np\right) }$ corresponding to the focus qudit $%
m$, and $\tau _{m\left( np\right) q}$, $\tau _{\left( mn\right) pq}$ are the
4-tangle of $\rho _{ABCD}$ corresponding to the foci $m$ and $\left(
mn\right) $ respectively, whilst the bracket is defined the same to that
above. From eq. (15), it is nicely seen that the right hand side of the
equation consists of the sum of the squared concurrence (the first term),
3-tangles (the second term) and 4-tangles (the last two terms). That is to
say the left hand side is the sum of different entanglement contributions.
It is obvious that some terms in the left hand side are repeated. One can
always eliminate the repeated ones by changing the factors before each term
in the right hand side. The reason is as follows. Take the term $C_{\left(
CD\right) (AB)}^{2}$ as an example. From eq. (8) and eq. (13), one can have
\begin{eqnarray}
C_{\left( CD\right) (AB)}^{2} &=&\frac{C_{\left( CD\right)
(AB)}^{2}+C_{\left( AB\right) (CD)}^{2}}{2} \\
&=&f(C_{mn}^{2})+F(\tau _{m\left( np\right) })+G(\tau _{\left( mn\right)
pq}),  \notag
\end{eqnarray}%
where $f(C_{mn}^{2})$ (squared concurrence), $F(\tau _{m\left( np\right) })$
(3-tangles) and $G(\tau _{\left( mn\right) pq})$ (4-tangles) are not
explicitly given. $C_{\left( CD\right) (AB)}^{2}$ can be eliminated by Eq.
(15) minus eq. (16). Analogously, the other repeated terms can be
eliminated. Hence, we define a quantity named global entanglement for the
given 4-partite pure state $\Psi _{ABCD}$ as%
\begin{eqnarray}
&&C(\Psi _{ABCD}) \\
&=&(C_{A(BCD)}^{2}+C_{B(ACD)}^{2}+C_{C(ABD)}^{2}+C_{D(BCA)}^{2}  \notag \\
&&+C_{\left( AB\right) (CD)}^{2}+C_{\left( AC\right) (BD)}^{2}+C_{\left(
AD\right) (BC)}^{2})^{1/2}.  \notag
\end{eqnarray}

It happens that every term in the right hand side of eq. (17) just
corresponds to the squared concurrence of the bipartite states generated by
bipartition of the given 4-partite pure state. In other words, so long as we
consider all the concurrences of the bipartite states after bipartite
partitions of the given 4-partite state, we can obtain the global
entanglement. In fact, this conclusion is not confined to the case of
4-partite systems. Following the above procedure, one can easily prove that
the analogous definition of the global entanglement for a given $n$-partite
state can be shown as the sum of all $m$-tangles with $m=2,3,\cdot \cdot
\cdot ,n$. Therefore, the global entanglement for any a state can be given
in the following rigorous way.

\textbf{Definition:-}If we consider the $i-to-(N-i)$ bipartite partitions of
an $N$-partite state $\left\vert \psi \right\rangle $, there exist $Num$
different bipartite states defined in $n_{1}(i)\times n_{2}(N-i)$ dimension,
where $Num=\left\{
\begin{array}{cc}
\sum\nolimits_{i=1}^{(N-2)/2}C_{N}^{i}+\frac{1}{2}C_{N}^{N/2}, & N\text{ \
is even} \\
\sum\nolimits_{i=1}^{(N-1)/2}C_{N}^{i}, & N\text{ \ is odd}%
\end{array}%
\right. $. The global entanglement $C(\left\vert \psi \right\rangle )$ can
be defined by
\begin{equation}
C(\psi )=\sqrt{\overset{P}{\underset{\alpha =1}{\sum }}\overset{Q}{\underset{%
\beta =1}{\sum }}\underset{p=1}{\overset{Num}{\sum }}\left\vert C_{\alpha
\beta }^{p}\right\vert ^{2}},
\end{equation}%
where $C_{\alpha \beta }^{p}=\left\langle \psi \right\vert S_{\alpha \beta
}^{p}\left\vert \psi ^{\ast }\right\rangle $ with $S_{\alpha \beta
}^{p}=L_{\alpha }\otimes L_{\beta }$; $L_{\alpha }$, $\alpha =1,$ $2$, $%
\cdot \cdot \cdot ,$ $P$ and $L_{\beta }$, $\beta =1,2,\cdot \cdot \cdot ,Q$
are the generators of $SO(n_{1})$ and $SO(n_{2})$, respectively, with $%
P=n_{1}^{p}\left( n_{1}^{p}-1\right) /2$ and $Q=n_{2}^{p}\left(
n_{2}^{p}-1\right) /2$; $p=1,2,\cdot \cdot \cdot ,Num$ denotes the $p$th
bipartite state.

It is obvious that a multipartite pure state $\left\vert \varphi
\right\rangle $ is fully separable if and only if $C(\left\vert \varphi
\right\rangle )=0$. The proof is omitted here.

In particular, when the definition is reduced to the tripartite quantum pure
states, $C(\psi )$ can be expressed by
\begin{equation}
C(\psi )=\sqrt{\overset{1}{\underset{\alpha =1}{\sum }}\overset{6}{\underset{%
\beta =1}{\sum }}\underset{p=1}{\overset{3}{\sum }}\left\vert C_{\alpha
\beta }^{p}\right\vert ^{2}},
\end{equation}%
where $C_{\alpha \beta }^{p}=\left\langle \psi \right\vert S_{\alpha \beta
}^{p}\left\vert \psi ^{\ast }\right\rangle $ with $S_{\alpha \beta
}^{1}=\sigma _{y}\otimes L_{\beta }$, $L_{\beta }$ are the generator of $%
SO(4)$; $S_{\alpha \beta }^{2}=L_{\beta }\otimes \sigma _{y}$ and $S_{\alpha
\beta }^{3}=\left( I\otimes swap\right) (L_{\beta }\otimes \sigma
_{y})\left( I\otimes swap\right) $ with $I=\left(
\begin{array}{cc}
1 & 0 \\
0 & 1%
\end{array}%
\right) $, $swap$ is the swap operator defined as $swap=\sum_{ikj^{\prime
}k^{\prime }}\delta _{jk^{\prime }}\delta _{j^{\prime }k}\left\vert
j\right\rangle \left\langle j^{\prime }\right\vert \otimes \left\vert
k\right\rangle \left\langle k^{\prime }\right\vert $, $j,k^{\prime
},j^{\prime },k=1,2$. Recalling the tensor cube in Ref. [19], one will find
that every $\left\vert C_{\alpha \beta }^{p}\right\vert $ just corresponds
to a plane of the cube including the surfaces and the diagonal planes (the
surfaces are corresponded to twice). However, unlike the criterion in Ref.
[19], the global entanglement has good properties which will be discussed in
the next section. But $3$ more complex optimal parameters have to be
introduced for the case of mixed states compared with Ref. [19].

In terms of the equivalent relations between the length of concurrence
vectors for bipartite pure states and the I-concurrence [5], we can rewrite
the global entanglement of multipartite pure states by%
\begin{equation*}
C(\left\vert \Psi ^{ABC\cdots N}\right\rangle )=\sqrt{\overset{P}{\underset{%
\alpha =1}{\sum }}\overset{Q}{\underset{\beta =1}{\sum }}\underset{p=1}{%
\overset{Num}{\sum }}\left\vert C_{\alpha \beta }^{p}\right\vert ^{2}}
\end{equation*}%
\begin{equation}
=\sqrt{2(Num-\sum_{p=1}^{Num}Tr\rho _{p}^{2})},
\end{equation}%
where $\rho _{p}$ denotes the reduced density matrix of bipartite pure
states corresponding to the $p-to-(N-p)$ partition of the given multipartite
pure state $\left\vert \Psi ^{ABC\cdots N}\right\rangle $. It is interesting
that eq. (20) is just equivalent to that in Ref. [20] in essence. Hence, the
global entanglement can also account for multi-partite correlations [20],
unlike Ref. [11,12]. In fact, from the viewpoint that global entanglement
consists of different entanglement contributions, the global entanglement
also consists of the sum of different correlations from physics, which
corresponds to full separability of a quantum state from mathematics.
Furthermore, we also show the connection with the distribution of
multipartite entanglement. In this sense, the global entanglement has its
own merit, even though a single measure can not sufficient to capture all
the properties of multipartite entanglement completely. In addition, one
should note the difference between ours and that in Ref. [20] ------Ref.
[20] considered all different reduced matrices, while we omit the repeated
bi-partitions.

\subsection{Extension to multipartite mixed states}

Analogous to Ref. [3,9], our global entanglement for multipartite pure
states can be extended to mixed states via minimizing their convex roofs,
\begin{equation}
C(\rho )=\inf \sum\limits_{k}\omega _{k}C(\psi ^{k}),
\end{equation}%
where the infimum is to be taken among all possible decompositions such that
\begin{equation}
\rho =\sum\limits_{k=1}\omega _{k}\left\vert \psi ^{k}\right\rangle
\left\langle \psi ^{k}\right\vert .
\end{equation}%
Considering $C(\psi )$ for pure states, $C(\rho )$ can be written by%
\begin{equation}
C(\rho )=\sum\limits_{k}\omega _{k}\sqrt{\overset{P}{\underset{\alpha =1}{%
\sum }}\overset{Q}{\underset{\beta =1}{\sum }}\underset{p=1}{\overset{Num}{%
\sum }}\left\vert \left\langle \psi ^{k}\right\vert S_{\alpha \beta
}^{p}\left\vert \psi ^{k\ast }\right\rangle \right\vert ^{2},}
\end{equation}%
where $P$, $Q$ and $Num$ are defined the same to the above section.

Following the analogous procedure [9], one can get
\begin{equation*}
C(\rho )=\sqrt{\overset{P}{\underset{\alpha =1}{\sum }}\overset{Q}{\underset{%
\beta =1}{\sum }}\underset{p=1}{\overset{Num}{\sum }}\left(
\sum\limits_{k}\left\vert (T^{T}A_{\alpha \beta }^{p}T)\right\vert
_{kk}\right) ^{2}}
\end{equation*}%
\begin{equation}
\geqslant \sum\limits_{k}\left\vert T^{T}\left( \overset{P}{\underset{\alpha
=1}{\sum }}\overset{Q}{\underset{\beta =1}{\sum }}\underset{p=1}{\overset{Num%
}{\sum }}z_{\alpha \beta }^{p}A_{\alpha \beta }^{p}\right) T\right\vert _{kk}
\end{equation}%
where $A_{\alpha \beta }^{p}=M^{1/2}\Phi ^{T}S_{\alpha \beta }^{p}\Phi
M^{1/2}$ and $z_{\alpha \beta }^{p}=y_{\alpha \beta }^{p}e^{i\phi _{\alpha
\beta }}$ with $y_{\alpha \beta }^{p}>0$, $\sum\limits_{\alpha \beta
p}\left( y_{\alpha \beta }^{p}\right) ^{2}=1$, the superscript $T$ denotes
the transpose of a matrix; Furthermore, we consider the matrix notation of
eq. (22) $\rho =\Psi W\Psi ^{\dag }$ and the eigenvalue decomposition $\rho
=\Phi M\Phi ^{\dag }$ and the relation $\Phi M^{1/2}T^{\dag }=\Psi W^{1/2}$.
Therefore, we obtain%
\begin{equation}
C(\rho )\geqslant \inf_{T}\sum\limits_{k}\left\vert T^{T}\left( \overset{P}{%
\underset{\alpha =1}{\sum }}\overset{Q}{\underset{\beta =1}{\sum }}\underset{%
p=1}{\overset{Num}{\sum }}z_{\alpha \beta }^{p}A_{\alpha \beta }^{p}\right)
T\right\vert _{kk}.
\end{equation}%
The infimum is given by $\underset{z\in C^{\alpha \beta p}}{max}\lambda
_{1}(z)-\sum\limits_{i>1}\lambda _{i}(z)$ with $\lambda _{j}(z)$ are the
singular values of $\overset{P}{\underset{\alpha =1}{\sum }}\overset{Q}{%
\underset{\beta =1}{\sum }}\underset{p=1}{\overset{Num}{\sum }}z_{\alpha
\beta }^{p}A_{\alpha \beta }^{p}$. Hence we get the lower bound%
\begin{equation}
C(\rho )\geqslant \underset{z\in C^{\alpha \beta p}}{max}\lambda
_{1}(z)-\sum\limits_{i>1}\lambda _{i}(z),
\end{equation}%
which is analogous to the result in Ref. [3,9]. It is obvious that eq. (26)
is the necessary condition for full separability.

As the applications to test the sufficiency, we evaluate $C(\rho )$ for
tripartite bound entangled states of qubits similar to Ref. [19]. For the
bound entangled state [13]
\begin{equation}
\bar{\rho}=\frac{1}{4}\left( 1-\sum_{i=1}^{4}\left\vert \psi
_{i}\right\rangle \left\langle \psi _{i}\right\vert \right) ,
\end{equation}%
where $\{\psi _{i}:i=1,\cdot \cdot \cdot ,4\}$ corresponds to $\{\left\vert
0,1,+\right\rangle ,\left\vert 1,+,0\right\rangle ,\left\vert
+,0,1\right\rangle ,\left\vert -,-,-\right\rangle \}$ with $\pm =(\left\vert
0\right\rangle \pm \left\vert 1\right\rangle )/\sqrt{2}$, one can get that $%
C(\bar{\rho})=0.1434$.

For D\"{u}r-Cirac-Tarrach states [14]
\begin{eqnarray}
\rho _{DCT} &=&\sum_{\sigma =\pm }\lambda _{0}^{\sigma }\left\vert \Psi
_{0}^{\sigma }\right\rangle \left\langle \Psi _{0}^{\sigma }\right\vert \\
&&+\sum_{k=01,10,11}\lambda _{k}\left( \left\vert \Psi _{k}^{+}\right\rangle
\left\langle \Psi _{k}^{+}\right\vert +\left\vert \Psi _{k}^{-}\right\rangle
\left\langle \Psi _{k}^{-}\right\vert \right) ,  \notag
\end{eqnarray}%
where $\left\vert \Psi _{k}^{\pm }\right\rangle =\frac{1}{\sqrt{2}}%
(\left\vert k_{1}k_{2}0\right\rangle \pm \left\vert \bar{k}_{1}\bar{k}%
_{2}1\right\rangle )$ with $k_{1}$ and $k_{2}$ the binary digits of $k$, and
$\bar{k}_{i}$ denoting the flipped $k_{i}$, Ref. [21] has shown that the
state is bound entangled for $\lambda _{0}^{+}=\frac{1}{3};\lambda
_{0}^{-}=\lambda _{10}=0;\lambda _{01}=\lambda _{11}=\frac{1}{6}$. In this
case, one can get $C(\rho _{DCT})$ =$0.2158$.

For the bound state [15]
\begin{eqnarray}
\rho _{bound} &=&\frac{1}{N}(2\left\vert GHZ\right\rangle \left\langle
GHZ\right\vert +\left\vert 001\right\rangle \left\langle 001\right\vert \\
&&+b\left\vert 010\right\rangle \left\langle 010\right\vert +c\left\vert
011\right\rangle \left\langle 011\right\vert +\frac{1}{c}\left\vert
100\right\rangle \left\langle 100\right\vert  \notag \\
&&+\frac{1}{b}\left\vert 101\right\rangle \left\langle 101\right\vert )+%
\frac{1}{a}\left\vert 110\right\rangle \left\langle 110\right\vert ),  \notag
\end{eqnarray}%
where $\left\vert GHZ\right\rangle =\left( \left\vert 000\right\rangle
+\left\vert 111\right\rangle \right) /\sqrt{2}$ and $N=2+a+b+c+1/a+1/b+1/c$,
one can easily obtain $nonzero$ $C\left( \rho _{bound}\right) $ for all $%
a=b=1/c$. The algorithm is the same to that in Ref. [19]. However we
conjecture that the sufficiency would be weaker for higher dimensional
systems.

\subsection{The properties of the entanglement measure}

Now we will show that the global entanglement is an entanglement monotone by
the similar method to that in Ref. [22]. From eq. (20), it is obvious that $%
C(\left\vert \Psi ^{ABC\cdots N}\right\rangle )=C(U_{i}\left\vert \Psi
^{ABC\cdots N}\right\rangle )$ where $U_{i}$ is a local unitary operation on
the $i$th subsystem. Furthermore, one can also find from eq. (20) that $%
C(\left\vert \Psi ^{ABC\cdots N}\right\rangle )$ is a concave function of
all the reduced density matrix $\rho _{p}$. At first, it should be noted
that the local operations are only performed on a subsystem and the
operations on one subsystem are independent of how to divide a multipartite
state into a bipartite one. E.g. we assume the subsystem $A$ is performed a
quantum operation $\{\varepsilon _{A,k}\}$ with $k$ denoting different
outcomes, then the final state after the operation is $\rho
_{f}=\sum_{k}p_{k}\rho _{k}$ where $p_{k}=Tr[\varepsilon _{A,k}(\left\vert
\Psi ^{ABC\cdots N}\right\rangle \left\langle \Psi ^{ABC\cdots N}\right\vert
)]$ and $\rho _{k}=(1/p_{k})\varepsilon _{A,k}(\left\vert \Psi ^{ABC\cdots
N}\right\rangle \left\langle \Psi ^{ABC\cdots N}\right\vert )$, which
directly implies that $\varepsilon _{A,k}$ is for the bipartition $A-others$%
. For the bipartition $AB-others$, $\varepsilon _{A,k}$ should be considered
as $\varepsilon _{A,k}\otimes I_{B}$ with $I_{B}$ the identity of subsystem $%
B$. Therefore, for different bi-partitions, $\varepsilon _{A,k}$ can always
be considered as the kronecker product of $\varepsilon _{A,k}$ and the
identities of other subsystems which are considered as a big subsystem of
the corresponding bipartite state.

For each $k$, let $\{r_{kl},\psi _{kl}\}$ be a pure-state ensemble realizing
$\rho _{k}$ optimally such that
\begin{equation}
C(\rho _{k})=\sum_{l}r_{kl}C(\psi _{kl}),
\end{equation}%
where $\sum_{l}r_{kl}=1$, $r_{kl}>0$ and $\rho _{k}=\sum_{l}r_{kl}\left\vert
\psi _{kl}\right\rangle \left\langle \psi _{kl}\right\vert $. Define $\sigma
_{kl}=Tr_{(A)}\left( \left\vert \psi _{kl}\right\rangle \left\langle \psi
_{kl}\right\vert \right) $ with the subscript $(A)$ denoting the big
subsystem of bipartite states which include subsystem $A$. Hence, due to the
concave $C(\rho _{p})$, we can get%
\begin{eqnarray}
C(\rho _{f}) &=&\sum_{k}p_{k}C(\rho _{k})=\sum_{kl}p_{k}r_{kl}C(\psi _{kl})
\notag \\
&=&\sum_{kl}p_{k}r_{kl}\sqrt{2(Num-\sum_{p_{r}=1}^{Num}Tr[(\sigma
_{kl})_{p}^{2}])}  \notag \\
&\leq &\sqrt{2[Num-\sum_{p_{r}=1}^{Num}Tr(\sum_{kl}p_{k}r_{kl}\sigma
_{kl})_{p}^{2}]},
\end{eqnarray}%
where $(\rho )_{p_{r}}$ denotes the reduced density matrix of $\rho $
corresponding to the $p-to-(N-p)$ partition of $\rho $. Note that for any $%
p-to-(N-p)$ partition of $\rho $
\begin{equation*}
\rho _{p}=Tr_{(A)}(\left\vert \Psi ^{ABC\cdots N}\right\rangle \left\langle
\Psi ^{ABC\cdots N}\right\vert )
\end{equation*}%
\begin{equation*}
=\sum_{k}p_{k}Tr_{(A)}(\rho _{k})=\sum_{k}p_{k}\sum_{l}r_{kl}Tr_{(A)}\left(
\left\vert \psi _{kl}\right\rangle \left\langle \psi _{kl}\right\vert \right)
\end{equation*}%
\begin{equation*}
=\sum_{kl}p_{k}r_{kl}\sigma _{kl},
\end{equation*}%
one can obtain that \
\begin{equation}
C(\rho _{f})\leq \sqrt{2(Num-\sum_{p_{r}=1}^{Num}Tr\rho _{p_{r}}^{2})}%
=C(\left\vert \Psi ^{ABC\cdots N}\right\rangle ).
\end{equation}

For the mixed state $\rho $, let $\{p_{k},\psi _{k}\}$ be a pure-state
ensemble realizing $\rho $ optimally such that $C(\rho )=\sum_{k}p_{k}C(\psi
_{k})$. Analogous to the pure-state case, considering a quantum operation $%
\{\varepsilon _{A,k}\}$, there exists the final state $\rho
_{f}=\sum_{k}p_{k}\rho _{fk}$ with $\rho _{fk}$ corresponding to every $\psi
_{k}$. According to eq. (24), $\ C(\psi _{k})\geq C(\rho _{fk})$ holds for
each $\psi _{k}$. It implies $\sum_{k}p_{k}C(\psi _{k})\geq
\sum_{k}p_{k}C(\rho _{fk})\geq C(\rho _{f})$, where the last inequality is
due to $C(\rho _{f})=\inf \sum\limits_{k}p_{k}C(\rho _{fk})$. All above show
that the global entanglement is decreasing under local quantum operations,
hence is an entanglement monotone.

\section{Conclusion and discussion}

We have presented the global entanglement for multipartite quantum systems
based on residual entanglement. Unlike the previous measure for multipartite
quantum states, the distinct characteristic of the global entanglement is
that the measure consists of the sum of different entanglement
contributions. Furthermore, we find that the global entanglement can be
conveniently obtained by the idea of bipartite partitions of a quantum
state. The measure has been shown to be an entanglement monotone. It is
interesting that the global entanglement for tripartite quantum pure states
of qubits has been effectively related to the tensor cube, and to that in
Ref. [20]. Hopefully the global entanglement can further our understanding
of multipartite entanglement.

\section{Acknowledgement}

This work was supported by the National Natural Science Foundation
of China, under Grant Nos. 10575017 and 60472017.


\begin{thebibliography}{99}
\bibitem{[18]} The equations are preserved only when the first terms in the
right hand side are different from each other.
\end{thebibliography}
\end{document}